# Electron quantum metamaterials in van der Waals heterostructures

Justin C.W. Song [1,2†] and Nathaniel M. Gabor [3,4,5]*

[1] Division of Physics and Applied Physics, School of Physical and Mathematical Sciences, Nanyang Technological University, Singapore 637371

[2] Institute of High Performance Computing, Agency for Science, Technology and Research, Singapore 138632

[3] Department of Physics and Astronomy, University of California, Riverside, California 92521, USA

[4] Laboratory of Quantum Materials Optoelectronics, University of California, Riverside, California 92521, USA

[5] Canadian Institute for Advanced Research, CIFAR Azrieli Global Scholar, MaRS Centre West Tower, 661 University Avenue, Toronto, Ontario ON M5G 1M1, Canada

† justinsong@ntu.edu.sg; * nathaniel.gabor@ucr.edu

**In recent decades, scientists have developed the means to engineer synthetic periodic arrays with feature sizes below the wavelength of light. When such features are appropriately structured, electromagnetic radiation can be manipulated in unusual ways, resulting in optical metamaterials whose function is directly controlled through nanoscale structure. Nature, too, has adopted such techniques – for example in the unique coloring of butterfly wings – to manipulate photons as they propagate through nanoscale periodic assemblies. In this Perspective, we highlight the intriguing potential of designer sub-*electron* wavelength (as well as wavelength-scale) structuring of electronic matter, which affords a new range of synthetic quantum metamaterials with unconventional responses. Driven by experimental developments in stacking atomically layered heterostructures – e.g., mechanical pick-up/transfer assembly – atomic scale registrations and structures can be readily tuned over distances smaller than characteristic electronic length-scales (such as electron wavelength, screening length, and electron mean free path). Yet electronic metamaterials promise far richer categories of behavior than those found in conventional optical metamaterial technologies. This is because unlike photons that scarcely interact with each other, electrons in subwavelength structured metamaterials are charged, and strongly interact. As a result, an enormous variety of emergent phenomena can be expected, and radically new classes of interacting quantum metamaterials designed.**

Electrons in solids have many different faces. For example, the diffraction of the electron wavefunction as it propagates through a material creates a complex pattern that determines its electronic properties (**Fig. 1a**). At the same time, an electron is a charged particle and possesses an electric field distribution that enables it to interact with other particles (**Fig. 1b**); this underlies its correlated behavior. In naturally occurring materials, these various (wavefunction-, Coulomb-) characteristics are *fixed* by its particular crystal structure and material environment. As a result, the *modus operandi* for the discovery of novel behavior in electronic systems often requires surmounting the challenges of growth and synthesis of completely new materials and systems.

An alternative, and rapidly progressing, nanotechnology is that of van der Waals (vdW) heterostructures[1], wherein stacks of atomically thin materials are assembled to form a complex quantum system. This has been driven from multiple angles including (*i*) the proliferation of available atomic layered materials[1-3] over a wide range of phases including semiconductors, semimetals, superconductors, ferromagnets, and topological insulators, (*ii*) means of assembling high quality vdW stacks (e.g., through mechanical pick-up/ transfer assembly[4,5]), as well as (*iii*) theoretical proposals of how these materials, when combined, can conspire to produce new physical phenomena. These two dimensional (2D) materials are a particularly attractive putty to use in molding materials-by-design since their electronic states are fully exposed and thus easily addressable. Few, if any, limitations are placed on the constituent layers since each layer is relatively stable in-plane[1]. Moreover, stacking such layers allows a far larger range of combinations than is conventionally possible via traditional epitaxial methods. All of these factors have led to a dizzying growth in the complexity of atomic layered stacks and diverse research avenues[6].

How should we view the prodigious progress in vdW heterostructures[1] to better navigate the path ahead? Here, we frame this rapid progress in terms of creating *quantum metamaterials* for electrons. As we opine below, atomic layered stacks provide powerful means of directly engineering the separate facets or character of electrons in solids, such as the texture of quasiparticle wavefunctions or the Coulomb fields a charged particle possesses (**Fig. 1**). This capability arises from the very small feature sizes (e.g., lateral feature size *a* and vertical distance *d*, see **Fig. 1c**) that are naturally formed in atomic layer stacks. Crucially, these features are tunable over multiple length scales that range from distances smaller than and up to the characteristic length-scale that defines each of the electron's aspects – e.g., the electron wavelength or unit cell size characterizes a quasiparticle's quantum wavefunction; screening length defines the scale over which Coulomb fields vary. Tuning these lateral and vertical features enables strong modification of electron behavior in loose analogy to optical metamaterials; electron quantum metamaterials we discuss below utilize both wavelength (or characteristic length-scale) as well as subwavelength scaled vdW structures. This is vividly illustrated in a prototypical stack of van der Waals materials (**Fig. 1c**) where a simple twist of the layer orientation allows nearly continuous variation of subwavelength (e.g., atomic-scale registration toggled by twist angle) as well as wavelength scale (e.g., superlattice unit cell size) features of the *electron*.

In the following, we lay out how vdW *structure* in electron quantum metamaterials with feature sizes at or below the various characteristic (electronic) length scales in vdW stacks can be used as design strategies for engineering quantum behavior. As we will see below, these strategies collectively fall into a length-scale engineering toolbox for emerging nanotechnologies - an intuitive framework for fundamental and technological progress.

**Atomic Layer Twists and Turns**

In mathematics, textiles, and art, intricate spatial interference configurations – the familiar moiré patterns – can be formed when two periodic templates are overlaid. Similarly, when two atomic 2D layers are stacked on top of each other, the periodic atomic structure within the constituent layers combines to create new spatial patterns **(Fig. 2)** that resemble artificial lateral superlattices. These naturally possess sub-*electron* wavelength features and enable a means of directly engineering the crystal unit cell (**Fig. 2**). When two 2D layers are commensurate[7-12], the electrons can be described exactly by new Bloch bands with an expanded unit cell defined by the superlattice structure (**Fig. 2a**). Even when the two layers have an incommensurate configuration, their moiré pattern[13] allows the low-energy quasiparticle excitations to be effectively described by a set of moiré superlattice Bloch minibands[7-11,14-16]. As such, we use here the same language of effective Bloch bands and effective unit cells to describe both cases.

**Engineering unit cell size.** Together with control over the unit cell, in particular its size, superlattices also possess an associated (reduced) Brillouin zone[7-11,14-16] that is tunable. The resultant Bloch mini-bands can be understood as bands folded in at the edges of the superlattice Brillouin zone. This vastly enriches the original bandstructure and grants control over the pattern of electron and hole *filling*[15,17-20] (the free carrier density and type) in the heterostructure. A striking example is the set of secondary Dirac cones formed from the moiré pattern in graphene/hexagonal Boron Nitride stacks[15,17-20] (G/hBN). In addition to a primary Dirac cone, replica Dirac particles are displaced at a higher energy defined by the superlattice, and exhibit tunable ambipolar behavior as the Fermi-energy passes through the secondary charge neutrality points[17-20]. Twisted bilayer graphene provides a particularly arresting illustration, where a rich array of minibands form and morph as a function of twist angle[7,9-11,21].

The facility to engineer unit cell sizes and its electron filling can lead to dramatic results when magnetic field is applied, for e.g., a fractal Landau level spectrum (Hoftstadter butterfly) can be observed[17-19]. The moiré superlattice sizes accessible in atomic layer heterostructures make such intricate behavior readily observable at realizable magnetic field strengths. This interplay between magnetic flux and superlattice size can even persist at high temperature where Brown-Zak magnetic oscillations have been induced[22] (**Fig. 2a**).

**Texturing the unit cell.** In optical metamaterials, sub-wavelength scale features alter the propagation of light. In electronic materials, structure that arises within the unit cell (the relative amplitudes and phases of the electron wavefunction as it passes through each of the atoms/orbitals of the crystal) provides a patterned texture to the periodic part of the Bloch wavefunction. This texture is responsible for quantum geometric (wavefunction) properties such as non-trivial pseudo-spinor texture, Berry curvature, electric polarization[23], and can mediate non-linear responses[24,25]. The expanded unit cells innate to layered heterostructures (**Fig. 2b**) provide a natural playground to design Bloch wavefunctions from the bottom up. In these, the specific atomic configuration can be tuned by twist angle, relative lattice constants, composition of each layer, as well as electric field effect.

*Unit cell structure and bandstructure.* Bilayer stacks with unit cells comprising atoms from both top and bottom layers are prime targets for texturing. A well-known example is (Bernal stacked) bilayer graphene, where the linearly dispersing Dirac dispersion in each of the constituent layers hybridizes to form a spectrum of massive charge carriers. Similarly, in trilayer stacks, the bandstructure morphs, possessing bands that can intimately depend on its stacking arrangement[26] (e.g., ABA vs ABC). In much the same way, the heterobilayer G/hBN develop band gaps near the Dirac point[19] due to AB sublattice asymmetry in the superlattice unit cell.

*Unit cell structure and quantum geometry.* Perhaps the most striking consequence of atomic layer twisting is that it enables us to produce new pseudo-spinor textures that are radically different from those found in the individual 2D layers. While the pseudo-spin of an electron wavefunction in graphene simply winds around the equator in a Bloch sphere, the pseudo-spin vector in G/hBN or biased bilayer graphene cants out-of-plane (**Fig. 2b**) and permits a non-zero Berry curvature to develop close to band edges[27]; Berry curvature encodes an intrinsic orbital angular momentum of wavepackets in the band structure induced by the configuration of atoms/orbitals in the unit cell[23].

Importantly, the patterned wavefunctions of twisted vdW heterostructures can provide a unique venue to control Berry phase effects that result from pseudo-spin texture[23]. Accessing the unusual quantum transport enabled by non-trivial wavefunction texture (quantum geometry) is currently under intense exploration in many material systems e.g., Weyl semimetals. Progress, however, is no more evident than in vdW heterostructures, where Berry curvature can mediate a wide variety of unconventional quantum geometric effects that include valley-selective Hall transport[27,28], non-local resistance[20,29,30], and valley (orbital) magnetic moments[27,31] to name a few. Quantum geometric effects can be particularly sensitive to the underlying symmetry of the crystal[24] wherein lowered crystal symmetry (e.g., engineered by strain) can lead to unusual effects such as a current induced (orbital) magnetization[32], or a non-linear Hall current[24] even at zero applied magnetic field.

*Unit cell structure, crystal symmetry, and expanded degrees of freedom.* The expanded unit cell in vdW stacks are also characterized by additional degrees of freedom (DOF). When top and bottom layers are symmetric, as in bilayer graphene, polarization on each of the layers acts as an additional DOF; breaking this symmetry either from an external electric control[33,34] or from spontaneous means[35,36] induces new phases[37]. The unit cell also embodies the symmetries of the larger heterostructure. For example, while monolayer $MoS_2$ breaks inversion symmetry, its bilayer $MoS_2$ preserves it; an additional out-of-plane electric field is required to break inversion symmetry/layer symmetry[31] and unlock Berry curvature of its the bands (**Fig. 2b**). This interplay of symmetry and the expanded DOFs becomes particularly striking when spin and valley (tied to magnetization) and layer (tied to electric polarization) DOFs are strongly coupled[38] enabling magneto-electric coupling and electrical access to inner DOFs.

**Topological bands.** Our ability to texture the Bloch wavefunctions in 2D twisted quantum metamaterials informs a unique possibility: designing topological bands out of trivial materials. Indeed, many two-dimensional materials (e.g. graphene, $MoS_2$, hBN, bilayer graphene) possess a Dirac-type electronic structure closely related to that found in topological insulators[39]. This "topological transmutation" can take one of several forms (**Fig. 2c**). For example, the stacking arrangement between top and bottom layers can slip at stacking faults. This sudden change in the crystal structure yields contrasting unit cell configurations across the fault line. In gapped bilayer graphene, such stacking faults host localized topological domain-wall states, which are gapless and valley-helical[40-42]; domain wall kink states can also be achieved by electric field using split-dual gates[43,44]. In 2D heterobilayers, manipulating the unit cell texture can enable the design of topological bands in commensurate G/hBN[45] and TMD bilayers[46]. Remarkably, when the atomic registration is incommensurate, one-dimensional (helical) kink states can proliferate and form a network-like structure (**Fig. 2c**) that can be switched on/off by electric field[46] and reconfigured spatially by strain.

Switchable topological bands, for which we can electrically turn on or off the helical edge states, also extends to intrinsic atomic layer topological insulators[47], e.g., that found recently in 1T'-$WTe_2$ [48-50]. The layered nature of vdW topological insulators means that they can be easily stacked, providing a straightforward means of parametrically increasing the edge conductance[47]; stacking can also enable to construct helical modes from combining edge states of opposite chirality in an electron-hole bilayer in the presence of a magnetic field[51].

**Quantum coloring: excited states.** While so far we have considered only the lowest energy excitations, the higher energy collective modes of quantum metamaterials can also be tuned by sub-*electron* wavelength features. Take for example an exciton, which is normally considered a simple (hydrogenic-like) bound state of an electron and hole. In conventional materials, the narrowly defined exciton energy gives rise to sharply resonant optical emission and absorption. In the presence of a unit cell structure with features smaller than the Bohr radius, the constituent electron and hole may possess Berry curvature. Because of this, the exciton optical spectrum morphs, exhibiting split angular momentum states[52,53]. The resultant change in the optical absorption is a unique quantum effect resulting from sub-electron wavelength features.

Similarly, plasmons - the collective modes formed from a high density of charge carriers - may experience chirality at zero-magnetic field[54,55] if the constituent electrons or holes possess non-trivial Berry curvature.

These are some examples of how collective modes may take on beyond-Landau Fermi liquid type characteristics[56] induced by non-trivial unit cells.

**Length scale engineering into the third dimension.** Beyond two layers, further stacking in the out-of-plane direction (three, four, and more layers) provides even more opportunity to engineer the structure and wavefunction texture of the unit cell. One interesting prospect is how motion in the out-of-plane direction can be coupled to motion in the in-plane directions. Inspiration may come from gyrotropic optical media, where left- and right-elliptical polarizations can propagate at different speeds through the material. Such handedness can, for example, be used to determine the chirality of molecules by measuring the change in linearly polarized light upon transmission. Carefully stacking and combining small atomic twists in a vdW heterostructure could realize such optical handedness in a 3D crystal structure[57] (**Fig. 2d**). While there are proposals for chiral materials composed of self-assembled inorganic materials[58], the electronic or optoelectronic behavior of such assemblies remains under investigation[59].

**Designer interactions in the solid state**

Unlike photons that travel independently through an optical medium, electrons may interact strongly in low dimensional materials. While physically confined to move within the 2D *x-y* plane, electrons in vdW materials are charged particles that possess electric fields extending out in all directions; in particular, the out-of-plane or *z*-direction. This is a direct result of the three-dimensional nature of the Coulomb interaction. Indeed, the extended *z*-direction profile of an electron's potential enables the electric field effect[60] - a ubiquitous means with which to control the carrier density in a 2D material by a proximal gate electrode. This character of the electron, its Coulomb fields (**Fig. 1b**), is responsible for a rich tapestry of phenomena - from scattering with phonons and impurities to bound electron-hole pairs and exotic correlated phases.

For vdW materials, vertical stacking (**Fig. 1c, Fig. 3**) enables the engineering of features that are far smaller than the characteristic length scales governing the electron's Coulomb fields. The exposed surfaces (and surface electronic states) of vdW materials make them particularly susceptible to the environment. Further, separate vdW layers can be stacked flush against each other or when electrical isolation of the layers is required, thin hBN spacer layers (as narrow as several atomic layers) can be used. This is in stark contrast to conventional bulk materials, where electrons residing deep in the bulk dominate material properties. As a result, the in-plane motion and dynamics of charges can be manipulated by their Coulomb fields out-of-plane.

**"Stray" fields and layer-to-layer interactions.** Coupling between in-plane DOFs with out-of-plane DOFs are an immediate way these "stray" Coulomb fields affect electronic behavior. A case-in-point is coupling of in-plane electrons with phonons in a nearby substrate or dielectric media (**Fig. 3a**). Phonons in the surrounding media, characterized by the displacement of lattice ions, possess dipoles that respond readily to extended electric fields of the electron. While present in all heterostructure stacks, the fully exposed electric potential profile of 2D layers enable taking this coupling to the extreme limit: for example, in insulator/graphene/insulator stacks, the scattering of electrons by phonons can be dominated by the surrounding dielectric. Indeed, G/hBN stacks have been shown to enable fast cooling of electrons in graphene by hyperbolic phonons in hexagonal boron nitride[61,62].

"Stray"-field coupling extends far beyond electron-phonon scattering. The out-of-plane extent of electric fields can mediate interaction between a wide range of DOFs and quasiparticles even when they are in separate planes (such as excitons, phonon-polaritons, and plasmons). For example, electrons propagating in one 2D layer may efficiently generate additional inter-layer electron-hole pairs[63] (**Fig. 3a bottom left**), rather than emit phonons. This multiplication process likely results from the increased interaction cross-section between electron Coulomb fields and the reduced phonon density of states in 2D materials.

Interestingly, layer indirect excitons that form in vdW material stacks exhibit a dipole moment that may be slightly canted away from the z-direction due to interlayer twisting[64,65]. The relaxation (or excitation) of such canted interlayer excitons involves a well-defined phonon mode. Such narrow phase space for exciton-phonon interactions could be used to engineer unusual devices that mimic vibronic transitions observed in photosynthetic energy transfer[66]. Coupling of this sort can even grant unconventional transduction between other DOFs; e.g., that of out-of-plane phonon-polaritons (charge neutral objects), which can be tuned electrically by a proximal metallic layer when they hybridize with plasmon-polaritons[67].

Perhaps even more striking is how charge DOFs residing in two nearby metallic layers separated by a 2D insulator interact via their Coulomb fields, even when there is no particle exchange between them. For instance, inducing a current in one 2D layer can drag carriers in the adjacent layer by exchange of momentum and energy[68]. When layer separation is smaller than the screening length, the strongly coupled regime (where inter-layer interaction is as strong as intra-layer interaction) can be accessed. For example, in this regime electrons and holes in the separated metallic layers condense to form an excitonic condensate[69,70] – a phase of matter distinct from that of its constituent parts (**Fig. 3a bottom right**). This is prototypical of an *interacting* quantum metamaterial and displays how the strong coupling of DOFs can lead to radically new behavior.

**Designer interactions: how can the Coulomb field be engineered?** Since field lines warp and contort depending on a vdW material's environment, stacking can provide the means to tailor the interaction between electrons. As an illustration, consider the simple structure of an atomically thin semiconductor (e.g., $MoS_2$) sandwiched on either side by insulators (**Fig. 3b**): dipolar fields between an electron and a hole in the same layer get squeezed due to the strong contrast of dielectric constant between the monolayer and its surroundings[71]. This warping of field lines leads to a non-hydrogenic series of excited states for 2D excitons and a dramatic modification of its band gap energy[72] that can be spatially engineered in atomically thin semiconductors (**Fig. 3b bottom**).

*Warping Coulomb fields by proximal gate.* Another simple structure consists of an atomically thin metal (e.g., graphene) placed very close to a bulk metallic plate (spaced by a 2D insulator). In such structures, the proximal bulk metal helps to screen electric fields caused by charge density fluctuations in the atomically thin metal. When positioned closer than a screening length apart, the effect can be significant. For example, a proximal graphite gate can smoothen out charge puddles in graphene leading to a more homogeneous carrier doping landscape[19,73]. Using similar approaches, plasmons in graphene/insulator/metal heterojunctions can be slowed down[74,75], and plasmon electric field profiles in the z-direction can be significantly squeezed into the plane[76] (**Fig. 3c**).

*Transforming Coulomb repulsion.* The warping and screening of Coulomb field lines can be taken to the ultimate limit: transforming the nature of the Coulomb interaction itself. One of the most remarkable proposals is to transmute the Coulomb repulsion (in a low-dimensional system) into an effective Coulomb *attraction* by placing it close to a highly polarizable medium[77]. In the early proposal by Little[77], electrons in the polarizable medium form a "glue" that mediates attraction in a low-dimensional conducting system; this can lead to a synthetic non-phonon mediated superconductivity. Strikingly, very recent evidence of *synthetic* electron attraction from Coulomb repulsion has been observed[78]. Since 2D conductors can be fully embedded in vertical stacks, yet have their field profiles fully exposed, vdW heterostructures can provide a unique platform to realize synthetic superconductivity[79,80].

**The tools of the trade in quantum metamaterials**. The designer-interactions-tool-kit described here outlines some emerging methods to control electron wavefunctions or the Coulomb fields of charge carriers. These tools principally exploit sub- and up-to- characteristic-length-scale features to engineer artificial quantum material behavior. Such tools, we believe, can be applied to other characteristic length scales and a wide range of as-yet-unexplored nanotechnologies. Given the myriad of correlated phases recently uncovered in 2D materials (e.g., gate tunable superconductivity in monolayer $MoS_2$[81], ferromagnetism in monolayer $Cr_2Ge_2Te_6$ [82] and $CrI_3$ [83], and anti-ferromagnetism in $CrI_3$ stacks[83]) the designer-interactions-tool-kit also provides new opportunities for tailoring and controlling correlated phases, particularly when distinct phases are proximal to each other[84-86]. Electric-field control is one of the most compelling attributes of utilizing vdW correlated phases since it can enable unusual couplings to arise: e.g., bilayer $CrI_3$ exhibits an electric-field switchable magnetic order[87,88].

We note that our design list is non-exhaustive: electrons also exhibit numerous other characteristics that can be directly engineered. A particularly conspicuous aspect is its spin DOF. This, too, can also be directly tailored by proximity coupling to a strong spin-orbit coupling layer[89,90], or a ferromagnet[91]. Another aspect is the vertical tunneling characteristic of stacks that can be tuned by layer arrangement[92,93] or magnetic ordering[94-97]. Stacks may also exhibit structural phase transitions when the twist angle is small[12] or have layer alignment that is dynamically controlled[98]; when coupled with electronic behavior these can lead to designer responses[98,99]. Lastly, we remark that stacking different vdW materials on top of each other is by no means the only way to create features in vdW heterostructures. For example, lateral superlattices can be engineered by patterning a vdW material with a local top-gate, or stacking a vdW material atop a pre-patterned dielectric substrate[100,101], or by strain (see e.g., Ref. 102). Patterned dielectric superlattices[101] are particularly attractive because they enable high quality lithographically defined superlattices possessing an electrically tunable mini-bandstructure[101] with the attendant properties of the superlattices discussed above; it also grants access to other types of structures (e.g., highly anisotropic one-dimensional superlattices[100]).

Our perspective collects early examples in this rapidly developing research direction and distills from them emerging design strategies for electronic control in vdW heterostructures, see Table 1 for the convenience of the reader. This sketches how each of the vdW heterostructure (experimental) tools (column 1) enable control over electronic properties (column 3) through overarching key design principles (column 2). The design principles act as quantum metamaterials strategies for vdW nanoengineers, and provide a rational path for tailoring materials properties. Given that many electronic properties (as well as strategies) are interlinked, one striking application of the design principles listed is using the principles in tandem. Such combination strategies can enable new functionality (see table 1), e.g., low unit cell symmetry can conspire with the field effect to grant electrical control over the internal quantum structure of electrons such as the spin structure (including out-of-plane component) and quantum geometry of TMDs, see Refs. 103-106 for a recent example in monolayer 1T'-$WTe_2$.

**Interacting quantum metamaterials in vdW heterostructures**

One of the most profound promises of quantum electronic metamaterials - and what sets them wholly apart from their photonic counterparts - occurs when electrons in subwavelength-structured samples interact to exhibit unexpected emergent behavior (**Fig. 4**). In such systems, interactions build on top of subwavelength features potentially providing a ripe venue for a variety of *complex* emergent phenomena that arise from relatively *simple* constituents. A simple illustration of the effect of cooperation between subwavelength features and interactions can be found in G/hBN (twisted) moiré superlattices. At the single particle level, the AB sublattice asymmetry (corresponding to an energy gap at the Dirac point) induced in graphene is expected to be minute due to the alternating gap sign in adjacent superlattice domains[14]. Yet due to the same superlattice structure, theory predicts that Coulomb interactions can renormalize the bare gap size with a faster scaling exponent (as compared to graphene without a superlattice structure) producing an immense enhancement to the gap size[107].

Perhaps the most compelling landscape for interacting quantum metamaterials is that of magic-angle twisted bilayer graphene heterostructures[108,109], where a set of new correlated phases have been found (**Fig. 4**). While strong coupling between top and bottom graphene layers at low twist angles produces a pattern of Bloch minibands, when the twist is close to "magic-angles" the electronic structure is pushed to an extreme limit[11,110]: nearly-flat electronic bands are formed. As a result, electron kinetic energy is quenched and the interaction parameter increases. At half-filling, strong correlations between electrons localized near AA stacking regions yield an insulating gap consistent with that of correlated insulator, such as a Mott insulator[108] (**Fig. 4b**). When doped away from half-filling, magic-angle twisted bilayer graphene superconducts[109]. This superconductivity exhibits doping-dependent domes reminiscent of high $T_c$ superconductors (**Fig. 4b**) and a critical temperature just under 10% of the Fermi temperature (indicative of strong coupling superconductivity).

What is the future of length-scale engineered nanotechnologies? The emergent behavior in twisted bilayer graphene was a surprise; it is a dramatic demonstration of how electron interactions and subwavelength features conspire in interacting quantum metamaterials to produce radically new phases. Strikingly, this strongly correlated behavior arose when two seemingly weakly correlated atomic layered materials (graphene) were stacked together – a quantum alchemy of sorts. Similar correlated insulating behavior has also been observed in ABC trilayer graphene on hBN heterostructures[111]. Together, these paint an exciting future ahead for nanotechnology based on van der Waals heterostructures. What other transmutations are possible when atomic layers, each possessing already ordered (interacting) phases are combined? The opportunities are vast.

**Acknowledgements**

The authors would like to thank Valla Fatemi, Frank Koppens, Paul McEuen, Javier Sanchez-Yamigishi, and Andrea Young for useful discussions, as well as Max Grossnickle from QMO Labs for graphics assistance. J.C.W.S. acknowledges support from the Singapore National Research Foundation (NRF) under NRF fellowship award NRF-NRFF2016-05 and a Nanyang Technological University (NTU) start-up grant (NTU-SUG). N.M.G. is supported by the Air Force Office of Scientific Research Young Investigator Program (YIP) award # FA9550-16-1-0216, and through support from the National Science Foundation Division of Materials Research CAREER award no. 1651247. N.M.G. also acknowledges support through a Cottrell Scholar Award, and through the Canadian Institute for Advanced Research (CIFAR) Azrieli Global Scholar Award.

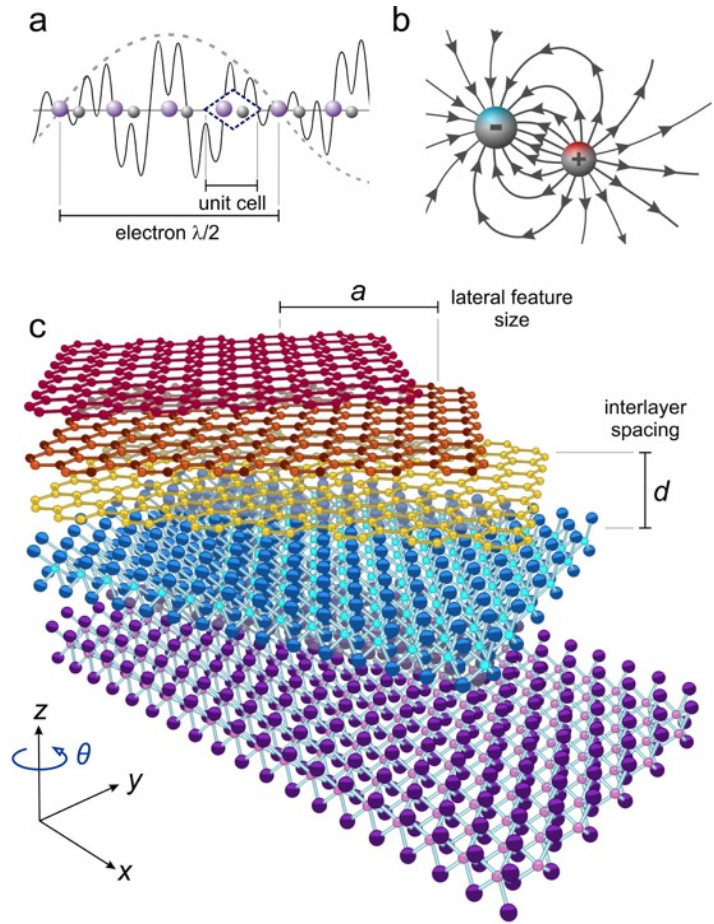

**Figure 1| Wavefunction- and Coulomb-characteristics of electrons in van der Waals heterostructure quantum metamaterials. a**, Schematic of the electronic wavefunction. The basic building block of the crystal, or crystal unit cell (outlined in dashed blue line), introduces repeating structure that modifies the electron matter wave of wavelength λ. **b**, Schematic of the electric field lines, a visualization of the local Coulomb field, surrounding an electron in proximity to a positive charge. **c**, a van der Waals heterostructure quantum metamaterial composed of individual 2D layers (transition metal dichalcogenides, graphene, and boron nitride) and characterized by lateral feature size *a*, interlayer spacing *d*, and atomic layer twist angle θ.

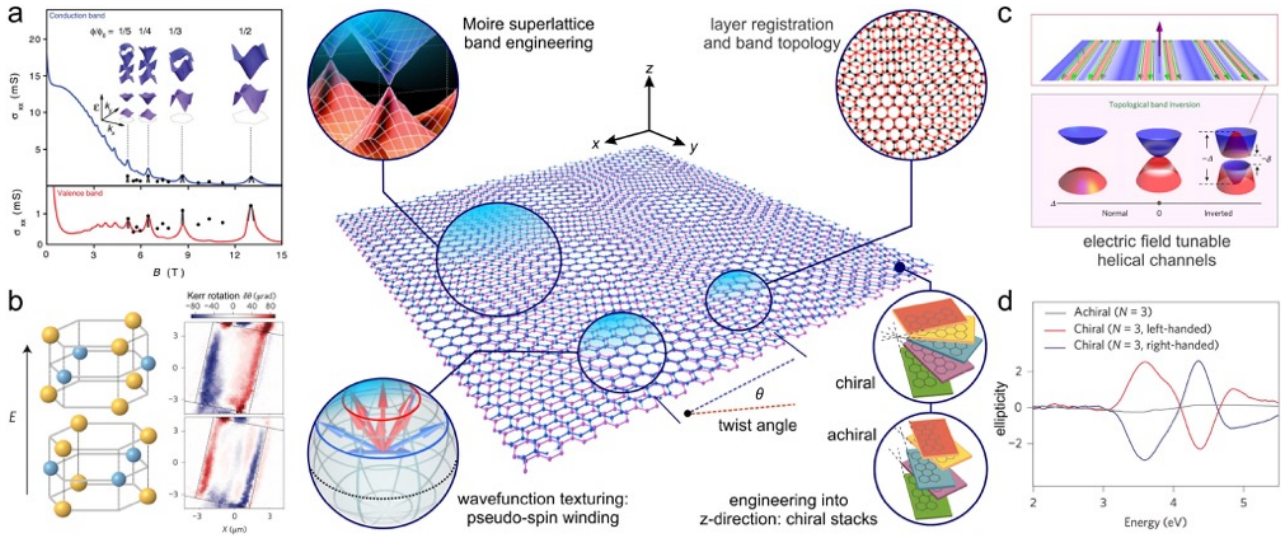

**Figure 2| Wavefunction engineering in van der Waals heterostructures. a**, (central) Expanded unit cell in moiré superlattices yield a new pattern of Bloch minibands. (side) Superlattice unit cell size, and magnetic flux can conspire to produce Brown-Zak magnetic oscillations in G/hBN superlattice that can persist to high temperature (adapted from Ref. 22). **b**, (central) The texture, comprising the relative phases and amplitudes of electron wavefunction on the orbitals/atoms in the stacked unit cell, can take on a non-trivial pseudo-spin winding. (side) Bloch band Berry curvature can be unlocked in bilayer $MoS_2$ by applying an out-of-plane electric field that breaks inversion symmetry. This manifests as a valley Hall effect, with valley density accumulating at sample edges producing a Kerr rotation (adapted from Ref. 31). **c**, (central) The local stacking configuration can influence the Bloch band topology. (side) When the local stacking registration changes spatially, one-dimensional (topological) kink states can manifest; in a TMD heterobilayer these helical kink states can be switched on/off by electric field (adapted from Ref. 46). **d**, (central) Stacking arrangement of vdW layers (e.g., handedness: chiral vs achiral) can program (side) ellipticity (circular dichroism) directly into the stack (adapted from Ref. 57).

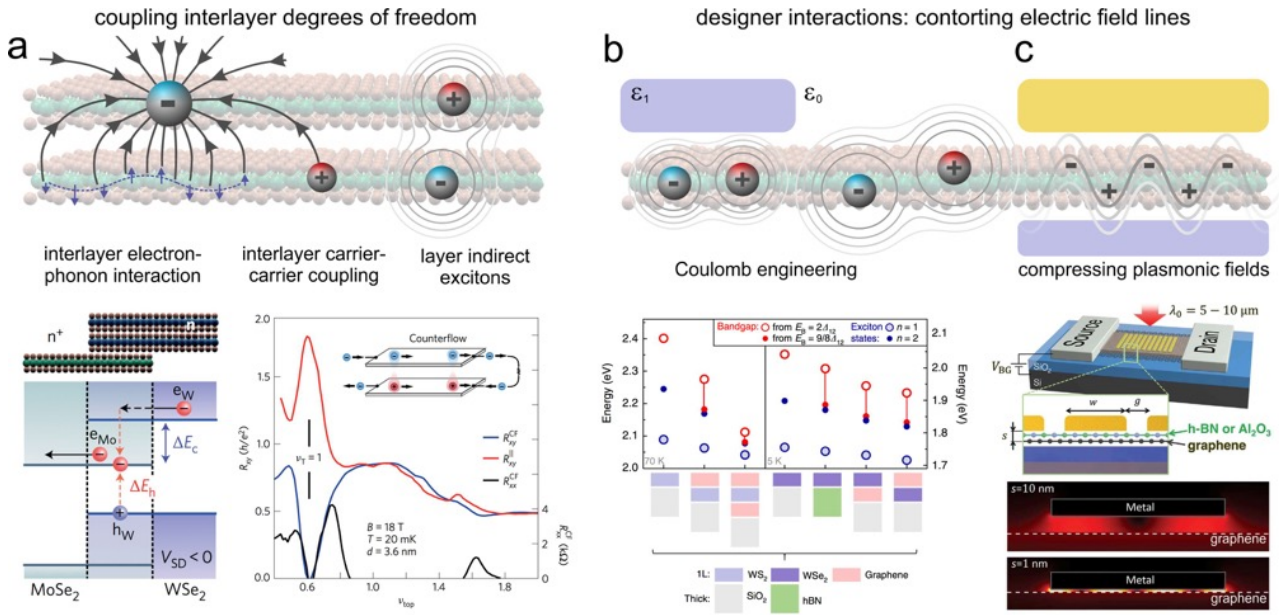

**Figure 3| Designing interactions in vertical van der Waals stacks. a**, (top) z-extent of Coulomb fields enable charge carriers in one-layer to couple/interact with other degrees of freedom in an adjacent layer. These DOFs include phonons, other charge carriers, excitons. (bottom left) Efficient inter-layer electron-hole multiplication generated from strong inter-layer carrier coupling (adapted from Ref. 63). (bottom right) Electrons and holes in separated vdW metallic layers condense to form an excitonic fluid characterized by a quantized Hall drag[69,70] (not shown), as well as a dissipationless nature (see counterflow geometry close to total filling = 1) in a double bilayer graphene heterostructure (adapted from Ref. 70). **b**, (top) Excitons possess Coulomb field lines that extend out of the 2D plane and are sensitive to environment adjacent to the 2D material. (bottom) These enable exciton spectra as well as the band gap to be modified by the dielectric environment (adapted from Ref. 72). **c**, (top) Similarly, the z-extent of a plasmon's electric field (as well as its in-plane charge distribution) is sensitive to the surrounding environment, allowing a nearby metallic plate to slow the plasmon down[74,75], or (bottom) compress the z-extent of its electric field[76] creating mode volumes up to $10^9$ times smaller than the free-space mode volume (adapted from Ref. 76).

**Table 1. Summary of key quantum metamaterial strategies for van der Waals heterostructures discussed in this article**

| vdW heterostructure tools | key concepts/strategies | electronic properties | examples of accessible phenomena |
|---|---|---|---|
| *lateral/vertical structure* atomic scale registration controlled by e.g., twist, stacking alignment, choice of homo/hetero-bilayers, strain. | wavefunction texturing | quantum geometry and Berry phase, unit cell symmetry, excited state internal structure, topological bands, internal magnetic degrees of freedom, chirality, bandwidth, layer degree of freedom | Valley Hall effect, magneto-electric effect, chiral Berry plasmons, topological bands from trivial materials, proximity induced spin-orbit coupling, stacking-tuned dichroism/tunneling, twist-tuned magic-angle flat bands, electrically controlled gap |
| | unit-cell size engineering | superlattice Bloch minibands, electron filling and density | secondary Dirac cones, Hofstadter butterfly |
| *vertical structure* proximal gate | gate/field effect | | |
| proximal dielectric stacks | stray fields and interlayer coupling | electron-phonon interaction, carrier-carrier scattering | hBN-hyperbolic phonon cooling, Coulomb drag |
| | Coulomb field engineering | field extent and profile, interaction sign | exciton/plasmon engineering, synthetic superconductivity |

---

**example of a combination strategy**

| | | | |
|---|---|---|---|
| | wavefunction texturing + gate/field effect | (low) unit cell symmetry, spin degree of freedom | electrical control over spin texture and Berry curvature in a monolayer TMD |

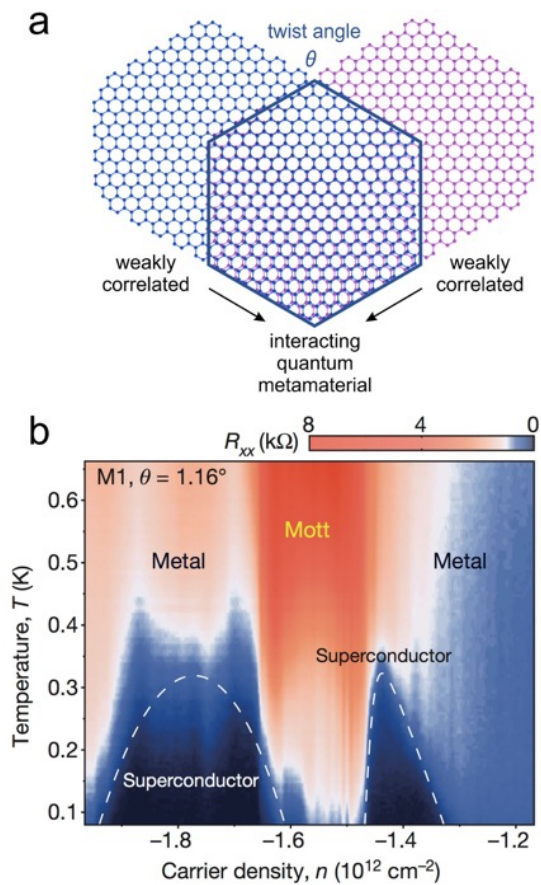

**Figure 4 | Strong correlations in interacting quantum metamaterials. a**, The lateral structure formed in vdW heterostructures can work synergistically with electron interactions to produce emergent correlated phases. **b**. Strongly correlated phases[108,109] have been found in twisted bilayer graphene when the twist angle between layers is close to certain "magic-angles". These phases can be readily tuned via gate from a correlated insulator (here it is labeled "Mott") to superconducting phase to a metallic phase (adapted from Ref. 109).